\documentclass{ceab}   

\usepackage{epsfig}     
\usepackage{graphicx}   

\usepackage{ceabbib}     
\usepackage[T1]{fontenc}

\begin{document}
\title{Classification of eclipsing binaries: attractive systems}

\author{O.~Yu. Malkov$^{1,2}$, E.~A. Avvakumova$^3$
\vspace{2mm}\\
\it $^1$ Institute of Astronomy of the Russian Academy of Sciences,\\
\it 48 Pyatnitskaya Street, Moscow 119017, Russia; malkov@inasan.ru,\\
\it $^2$ Faculty of Physics, Moscow State University, Moscow 119992, Russia,\\
\it $^3$  Department of Astronomy and Geodesy,
\it Ural Federal University,\\
\it 51 Lenin Street, Yekaterinburg 620000, Russia;
\it Ekaterina.Avvakumova@usu.ru
}

\maketitle

\begin{abstract}
We have compiled a catalogue of eclipsing variable stars,
the largest catalogue, containing classified eclipsing binaries.
A procedure for the classification of eclipsing binaries,
based on the catalogued
data, is also developed. It was applied to unclassified eclipsing binaries.
In this paper we discuss eclipsing binaries, which can not be
classified with the procedure. Some of them belong to
marginal evolutionary classes. Observational data for others are
too contradictory, and additional observations are needed to
attribute them to one or another evolutionary class.
\end{abstract}
 

\section{Introduction}
Eclipsing binaries represent one of the most numerous type of
binaries and serve as an invaluable source for determination
of the fundamental
properties of stars: masses, radii, temperatures and luminosities. 

\begin{table}
\setlength{\tabcolsep}{2pt}
\begin{tiny}
\caption{Catalogue of eclipsing variables, main table, first fifteen lines.}\label{tab:catalogues}
\begin{tabular}{llllrrrlrlrrrrrll}
\hline
(1)    & (2)& (3)& (4)& (5)   & (6)  & (7)  & (8)&      (9)      &(10)& (11)& (12)& (13)& (14)& (15)& (16)    & (17) \\[1mm]
\hline
RT And & CB & EA & RS &  8.97 & 0.41 & 0.31 &  V &  0.6289 &  v & 170 &   0 & 170 &   0 &     &  F8V+K1 & a \\
SY And &    & EA &    & 10.70 & 1.50 & 0.00 &  V & 34.9085 &    &  60 &  27 &     &     &     &  A0+K1  &   \\ 
TT And & SA & EA &    & 11.50 & 1.50 & 0.10 &  V &  2.7651 &  v & 140 &   0 &     &     &     &  A+G7IV &   \\
TW And & SA & EA &    &  8.98 & 2.06 & 0.15 &  V &  4.1228 &  v & 130 &  20 &     &     &     &  F0V+K0 &   \\
DK And &    & EW &    & 12.50 & 0.60 & 0.60 &  p &  0.4892 &    &     &   0 &     &   0 &     &  -----  &   \\
UU And & SA & EA &    & 11.20 & 3.00 & 0.20 &  V &  1.4863 &  v & 170 &   0 &     &     &     &  A8IV/V &   \\
WW And & SA & EA &    & 10.92 & 0.67 & 0.16 &  V & 23.2852 &    &  50 &  12 &     &     &     &  A0:+G5-K0III &   \\
WX And &    & EA &    & 12.10 & 1.70 & 0.00 &  V &  3.0011 &  d & 120 &  30 &     &     &     &  F5IV   &   \\
WZ And & CB & EB &    & 11.60 & 0.40 & 0.01 &  V &  0.6957 &  v &     &   0 &     &   0 &     &  F5     &   \\
XZ And & SA & EA &    &  9.91 & 2.54 & 0.25 &  V &  1.3573 &  v & 160 &   0 & 260 &     &     &  A1V+G5IV &   \\
AA And & CB & EA &    & 10.30 & 0.90 & 0.30 &  p &  0.9351 &    & 210 &   0 &     &     &     &  B8V      &   \\
AB And & CWW& EW &    &  9.49 & 0.97 & 0.83 &  V &  0.3319 &  v &     &   0 &     &   0 &     &  G5+G5V   &   \\
AD And &    & EB &    & 11.20 & 0.62 & 0.58 &  V &  0.9862 &  v &     &   0 &     &   0 & 494 &  A0V      &   \\
AM And &    & EA &    & 12.50 & 1.20 & 0.00 &  p &  8.8505 &    &  80 &     &     &     &     &  -----    &   \\
AN And & DM & EB &    &  6.00 & 0.16 & 0.09 &  p &  3.2196 &    &     &   0 &     &   0 &     &  A7Vm     &   \\[1mm]
\hline
\multicolumn{17}{p{12cm}}{{\bf Identification, classification}: 
(1) name of the star (as listed in GCVS); 
(2) class of the system (the abbreviations are given according to sections 2.2.1--2.2.3); 
(3) morphological type of the light curve (EA, EB, EW; as in the GCVS); 
(4) additional variability type from the GCVS. {\bf Photometry}: 
(5) magnitude at maximum brightness; 
(6) depth of primary minimum, $A_1$, mag; 
(7) depth of secondary minimum, $A_2$, mag; 
(8) the photometric system for magnitudes as in the GCVS. 
{\bf Period}: 
(9) period of the variable star, $P$, day; 
(10) information on the sign of period variability 
(d: derivative is negative and period decreases, 
i: derivative is positive and period increases, 
v: derivative is non-zero and the sign varies 
thus period increases and decreases, 
u: derivative is non-zero, but the sign is unknown 
thus period increases or/and decreases). 
{\bf Eclipses} (in phase $\times$ 1000 unit): 
(11) duration of primary eclipse, DI; 
(12) duration of totality in primary eclipse, dI; 
(13) duration of secondary eclipse, DII; 
(14) duration of totality in secondary eclipse, dII; 
(15) phase of secondary minimum, MinII-MinI. {\bf Spectra}: 
(16) spectral types and luminosity classes of the primary 
and secondary; 
(17) information on chromospherical activity 
(a: chromospherically active system).}
\end{tabular}
\end{tiny}
\end{table}

%
Catalogue of Eclipsing Variables (CEV) was compiled by \citet{CEV}.
CEV has been used to develop the most comprehensive set of rules 
for the classification of eclipsing binaries to date \citep{Malkov2007}.

For compilation of the first version of catalogue \citep[see][sec. 2.1]{Malkov2007}
we used data from General Catalogue of Variable Stars, GCVS (version 2004) 
and its textual remarks \citep{gcvs}. A lot of new eclipsing variables were 
discovered since 2004, and currently GCVS contains about 7000 
eclipsing binaries. We have added the new stars in the second 
version of CEV.  
The current version of CEV contains data for 7179 eclipsing binaries 
and represents 
the largest list to date of eclipsing binaries classified from 
observations.
Catalogue is described in detail by \citet{Avvakumova2012}. 
A lot of additional bibliographic sources were used to update data in CEV. 
The list of the sources with the comments is also given in the \citet{Avvakumova2012}.
Altogether information on 1783 binaries was updated.
First fifteen lines of the catalogue are presented
in table~\ref{tab:catalogues}.
The catalogue will be available in electronic form at the CDS
via anonymous ftp to
cdsarc.u-strasbg.fr or via http://vizier.u-strasbg.fr/viz-bin/VizieR

We have investigated the distribution 
of the catalogued systems in various observational planes 
and extracted from them a number of rules that allows us
to make classification of catalogued systems based on a set of observational 
parameters, even if the set is incomplete. 

When procedure of classification has been applied to previously
unclassified CEV systems, we have found a 
number of eclipsing binaries, which can not be classified
with the procedure.
Some of them, apparently, belong to marginal
evolutionary classes. Observational data for others are too
contradictory, and additional observations are needed to
attribute them to one or another class.

The scheme of eclipsing binaries classification,
adopted in the 
present study, is described briefly in Section~\ref{ev}
Our method of eclipsing binaries classification, based on
catalogued data, and principal results,
including unclassified binaries, are discussed in Section~\ref{class}
Finally, in Section~\ref{conclusions}, we draw conclusions.

%


\section{Evolutionary classes of eclipsing binaries}\label{ev}
The catalogue CEV supplies data with independently determined evolutionary
classes of systems. The list of the original 
catalogues and data sources we used have been listed in 
\citet{Avvakumova2012}.

According to well known schemes, we have divided all 
binaries into three classes: detached, semidetached 
and contact (denoted as D, S, C, respectively) systems.
Moreover, in each of these 
classes, sub-classes can be distinguished. %


%
\subsection{Detached systems}
\textbf{Main sequence systems (DM)}\\
Both components are main sequence stars which do not fill 
their inner Roche lobes. 

\noindent\textbf{Systems with two subgiants (DR)}\\
In these systems,
both stars are subgiants, they do not fill their Roche lobes 
(alternative names: AR Lac systems, RS CVn systems or long-period 
RS CVn systems). 
The hotter component in such systems is usually less massive 
and smaller and has a spectral type of either F or G. 

\noindent\textbf{Giant and supergiant systems (DG)}\\
Based on value of period and spectral type
we have devided DG systems on two subclasses: E (early-type)
and L (late-type). Binaries of the E-subclass systems consist 
of two hot stars: well evolved primary 
(WR, giant or supergiant) and MS early 
type or hot giant secondary. Orbital 
periods of DGE systems are shorter than 35 days. 
L-subclass systems include two stars with large 
temperature difference. 
Primaries of such systems are
MS, giant or supergiant stars of spectral type
from late-B to mid-F while secondaries  are
always late type (from late-G to M) evolved stars.
Orbital periods of L-subclass systems are about 100 
days and longer. 


%
\noindent\textbf{White dwarf systems (DW)}\\
Detached systems with white dwarf (or subdwarf) as a primary. 
The secondary usually is a low-mass star (main sequence or subgiant).
%

\noindent\textbf{Symbiotic detached systems (D2S)}\\
According to \cite{Belczynski2000} there are five detached 
symbiotic binaries which comprise a red giant 
transferring material to a white dwarf via a stellar wind.

%
\subsection{Semidetached systems}
\noindent\textbf{Classical algols (SA)}\\
Here the more massive 
component lies in the range from middle B to early F, and the 
other component is of type F or later. The primary (hotter, brighter 
and more massive) component is assumed to have a normal mass, radius and luminosity
for its spectral type. Also the 
primary can have larger (if it is a donor) or smaller (if it 
is an accretor) radius than the secondary. 
The secondary (cooler, overluminous and oversized) component 
is assumed to fill its Roche lobe. 

%
\noindent\textbf{Cool semidetached systems (SC)}\\
Based on 
the definition of \citet{Popper1980}, both components are late 
type subgiants or giants.

\noindent\textbf{Hot semidetached systems (SH)}\\
According to \citet{Popper1980} 
the hotter component of these systems is an early B-type star and 
the cooler is of type B or an early A.

\noindent\textbf{Cataclysmic semidetached systems (S2C)}\\
These systems 
contain a white dwarf (or a white dwarf precursor) primary and a 
low-mass secondary, which fills its critical Roche lobe. The secondary 
is not necessarily unevolved. 
We use catalogues of 
\citet{Ritter2003} and \citet{Downes2001} to classify this type 
of objects.

\noindent\textbf{X-ray semidetached systems (S2H, S2L)}\\
According to \citet{Liu2006} and \citet{Ritter2003} there are 
two sub-classes of X-ray semidetached binaries: massive
(denoted as S2H) and low-mass (denoted as S2L) systems. 
Massive X-ray binaries comprise a compact object
orbiting a massive OB star (supergiant or Be star).
Low-mass X-ray systems  consist of either a neutron star or a 
black hole primary  
and a low-mass secondary which fills its critical Roche lobe. 
%
%
\subsection{Contact systems}
According to \citet{Wilson1979}, these systems are all in fact overcontact 
systems with both components having surfaces larger than their 
critical Roche lobes. They are synchronous, circular orbit systems 
with common envelope. Besides overcontact systems, in this section 
we also consider near-contact systems.

\noindent\textbf{Near-contact systems (CB)}\\
\citet{Pribulla2003} designate near-contact binaries as B-type 
systems whose components are in physical but not in 
thermal contact. They usually consist of two stars of very 
different effective temperature enclosed in a common envelope.
According to \cite{Shaw1994} we distinguish F-type (CBF, FO Vir 
is a prototype)  and V-type (CBV, V1010 Oph is a prototype) near-contact
binaries. In V1010 Oph type systems the primary component (hotter) 
is at or near the Roche lobe, while the secondary (cooler) is inside 
the critical lobe and light curves are usually asymmetric.
In FO Vir type binaries, 
the secondary is at or near its Roche lobe and the primary is inside 
the Roche lobe.
Light curves for these systems are always symmetric. In both cases the
primary has normal size but the secondary is oversized.

\noindent\textbf{Early-type systems (CE)}\\
They are contact systems 
of early spectra (not later 
than A0 )with both components close to their Roche lobes. 

\noindent\textbf{Late-type systems (CW)}\\
They are contact systems 
with spectrum of primary usually later than about A-F (also 
known as W UMa systems). 
According to \citet{Binnendijk1977}, classical CW 
systems were divided into A-type (CWA, the larger star 
is hotter and primary minimum is the transit) and W-type  
(CWW, the smaller star is hotter, the primary minimum is 
the occultation).

\noindent\textbf{Giant systems (CG)}\\
Both components of these 
systems are early-type very luminous giants or 
supergiants close to their critical lobes. Orbital periods 
of CG systems are longer than those of CE binaries.

\section{Classification of eclipsing variables}\label{class}
\begin{figure}
\begin{center}
\epsfig{file=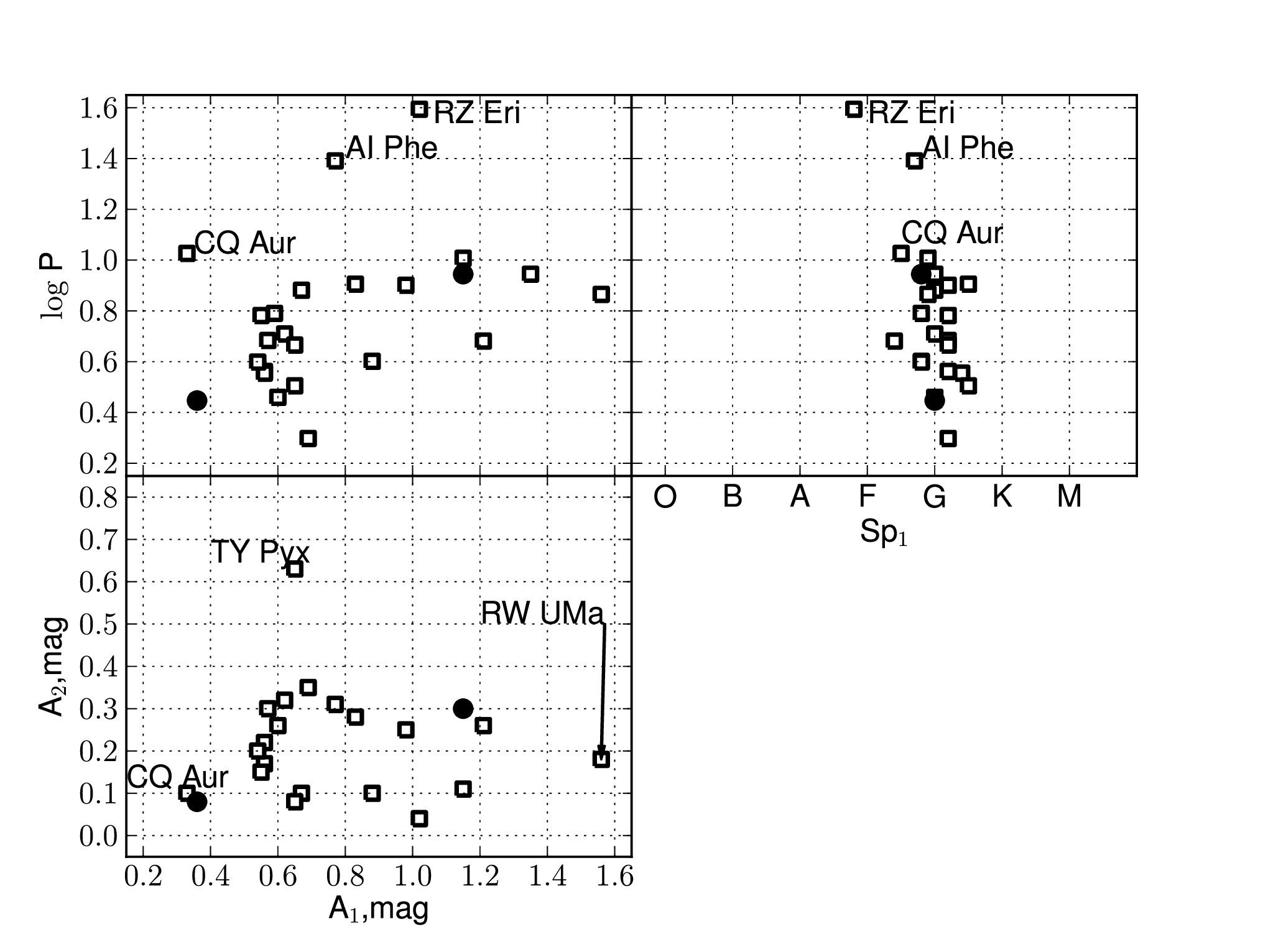,width=14cm}
\end{center}
\caption{The distribution of detached systems with subgiants (DR systems)
on A$_1$ -- A$_2$ -- $\log$P -- Sp$_1$ planes.
Catalogued DR systems are 
shown with squares. Filled circles denote two systems which have 
been classified as DR with our classification algorithm
(see text for details).}\label{fig:DRclass}
\end{figure}

To develop the method of classification we have performed analysis of the 
distribution of observable stellar parameters of eclipsing systems 
on various diagrams such as (A$_1$ vs A$_2$), (A$_1$ vs $\log$P), 
(Sp$_1$ vs Sp$_2$) etc. An example of such analysis for 
detached 
systems with subgiants (denoted as DR in the catalogue, most of them 
belongs to RS CVn systems) is shown in Fig.~\ref{fig:DRclass}.

As can be seen from the figure (bottom panel), the depth of primary minimum A$_1$ 
for the most DR systems lies between
0.5 mag and 1.2 mag, while the depth of secondary minimum A$_2$ is not greater than 0.6 mag.
Three exceptional cases are CQ~Aur (A$_1$=0.33 mag),
RW~UMa (A$_1$=1.56 mag) and TY~Pyx (A$_2>$ 0.6 mag).
Data for one of them, RW~UMa, are obsolete and
the system needs a further study.

Most of DR systems 
have value of period (left top panel of Fig.~\ref{fig:DRclass}) less than 10 days, except of RZ~Eri and AI~Phe. 
Both systems are long-period RS~CVn systems. Lower limit for period 
value equals to about 1.5 days.

The spectral 
type of the hotter component (on the right top panel of Fig.~\ref{fig:DRclass}) of the most of catalogued
DR systems is F-G with luminosity class IV or V.
Only RZ Eri with spectral
type of the hotter component A8-F0IV does not satisfy this rule.
We have found also that secondary spectrum of such systems
are G to mid-K with luminosity class IV or V.
All of DR systems are 
chromospherically active stars.

Also, all of DR systems exhibit EA-type light curves 
(i.~e. the light of binary remains almost constant between eclipses).

We have used all conditions listed above to classify systems with unknown 
evolutionary type in our catalogue and have found two
additional detached systems
with subgiants. These two systems are shown with filled circles in 
Fig.~\ref{fig:DRclass}. We have examined these systems with SIMBAD. One of them, RS~Ari,
is included in the catalogue of \citet{Brancewicz1980} as a detached 
system and in the catalogue of chromospherically active binaries of 
\citet{Eker2008}. The other system, CF~Tuc, is a detached and well studied 
RS CVn binary star \citep{Dogru2009}. So our classification of these
two binaries as detached systems with subgiants is confirmed.

After comprehensive analysis of our catalogue we have developed similar 
sets of rules for the classification of other types of binaries in our 
catalogue.

The full set of rules can be applied only for systems with all known
parameters, i.~e. amplitudes, periods and spectra. 
In this case our procedure has allowed us to classify about 77\% of systems.
Therefore there is a number of systems in CEV that we could not
classify for some reason. Three main reasons are discussed below. 

\subsection{Lack of observations}
In the case of incomplete set of parameters (for example period
or/and spectra are unknown) a lot of systems have been classified ambiguously. 

\begin{figure}
\begin{center}
\epsfig{file=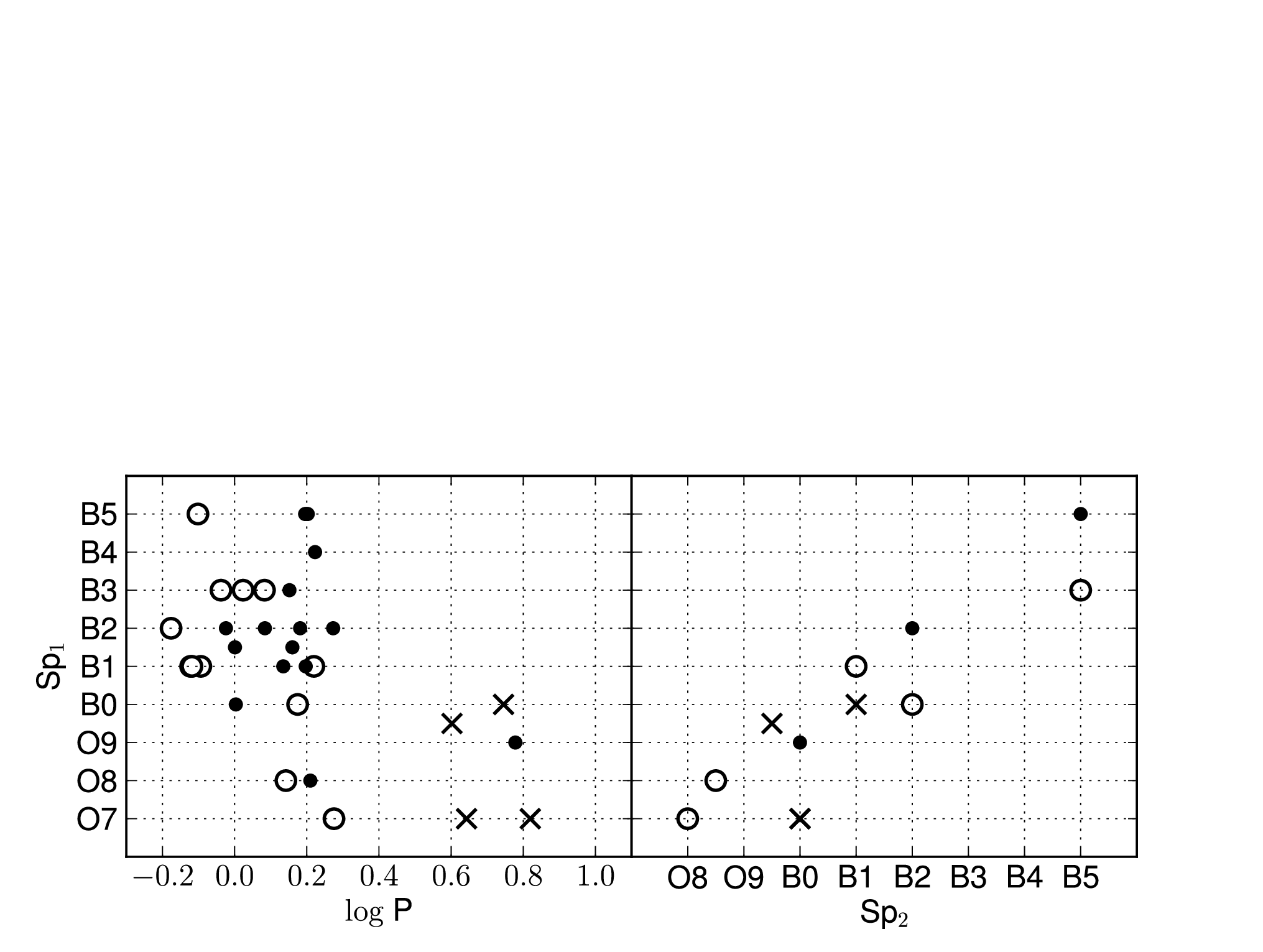,width=14cm}
\end{center}
\caption{The distribution of contact early-type systems (CE and CG systems)
on $\log$P -- Sp$_1$ --  Sp$_2$ planes.
Catalogued CE systems are 
shown with empty circles, catalogued CG systems are shown with crosses,
filled circles denote ``candidates'' to CE or CG class
(see text for details).}\label{fig:CECG}
\end{figure} 

For example we have found a number of ``candidates'' to CE or CG classes,
and they are shown on Fig.~\ref{fig:CECG}.
On the left panel of Fig.~\ref{fig:CECG}, the distribution of classified
CE (contact early type binaries) and CG (giant and supergiant contact
binaries) on the plane ``primary spectrum --- period'' is shown
with circles and crosses, respectively. On the right panel, classified CE
and CG systems on the plane ``primary spectrum --- secondary
spectrum'' are shown. The ``candidates'' to both CE and CG classes
are shown with filled circles. Information about secondary spectrum is
available only for three ``candidates''. As one can see, CG systems have
longer periods as compared with CE and it will be possible to divide CE
binaries and CG binaries if only primary (hotter) spectrum and period
are known. But without secondary spectrum all these ``candidates'' may
be classified as DM (detached main sequence), DG (giant and supergiant
systems) or SH (hot semidetached algols)! So information about secondary
spectrum is necessary for a lot of systems in CEV to classify them
properly.

Another problem that we can refer as ``lack of observations''
is an old photographic data about amplitudes of primary and secondary
minima. One such system RW~UMa have been mentioned earlier (see
Fig.~\ref{fig:DRclass}).
RW~UMa were classified as DR by
\citet{Popper1990} and values of A$_2$, period and spectra are
normal as compared to other DR binaries. But value of depth of
the primary minima is equal to $1.56^m$ and is larger than typical
values for other DR.  It may be assumed that new photometric
observations will help us to determine new value of A$_1$ and
it will be lower than $1.2^m$.

While
updating the catalogue we have found a number of systems, whose
parameters contradict each other or/and too unusual for
the system's evolutionary class.
In the majority of cases such parameters are based on photographic 
(probably,
outdated) photometry, and we could not find confirmation
(or refutation) of that data in the literature.
Consequently, new observations and further investigations
are required for these systems.
The list of these systems contains 238 eclipsing binaries.
In particular, it contains
systems with too large depth of primary minimum $A_1$.
It also contains contact and semi-detached systems, exhibiting
evidence of orbital eccentricity
(which is strange for such close systems): phase of secondary minimum
differs from 0.5 or duration of primary and secondary eclipses
are not equal.

\subsection{Contradictory classification}
One of the result of our classification is that a number of systems
have turned out to belong simultaneously to several classes.
Observed parameters (periods, spectra, depth of minima, etc) are
inconsistent with each other. All of such systems were checked with
the literature. For a lot of them we have found contradictory
classifications. Several examples are listed below.

\noindent\textbf{RT~Lac}\\
This system is one of the most ``peculiar'' in our list. Its period
is as for classical semidetached algols (denoted as SA) and detached
systems with subgiants (DR), but spectrum of the hotter primary
component corresponds to typical spectra of cool algols (SC in our
notation) and DR.  Additionally  we can use the relation between
spectra of primary and secondary. Such a relation for RT~Lac points
out to SC and DR classes. Moreover the depth of the secondary
is greater than for other SA, SC and DR systems! We could include
it in the DR class because of its RS~CVn type of variability
\citep{Strassmeier1993}, but most of DR type systems have depth
of secondary minima A$_2$ not greater than $0.4^m$, while A$_2$
of RT~Lac equals to $0.83^m$. According to \citet{Ibanoglu2001}
the brightness of the system at three phases, i.e.,
mid-primary and quadratures, shows quasi-periodic changes.  The
brightness at the primary eclipse (phase 0.0) shows the largest
variation with a maximum amplitude of about 0.3 mag in the B and V
filter. The light variations at the secondary maximum (phase 0.75) resemble
those at primary eclipse but with a maximum amplitude of about 0.2
mag, while the variations at primary maximum (phase 0.25) are generally
in the opposite sense, but with a lower amplitude (about 0.1 mag). Thus the significant out-of-eclipse variations may
lead to a large A$_2$ value. Again \citet{Ibanoglu1997s} have pointed out to semidetached configuration of RT~Lac, so why it is not SA or SC? As it was mentioned above we could not classify this system as SA because of its spectra, and its period is some lower than typical values for SC systems.

\noindent\textbf{AO~Cas}\\
AO Cas has been classified as contact early type binary (CE) according
to \citet{Polushina2004}, but its period ($3^d.52$) is larger than
upper limit for CE systems ($1^d.89$). Such a large period is typical
for contact systems with giants and supergiant (denoted as CG in CEV),
but for this class the lower limit for A$_1$ equals to $0.32^m$
while A$_1$ for AO~Cas is $0.17^m$. According to \citet{Gies1991}
and \citet{Bagnuolo1991} the system is semidetached,
less massive secondary is subgiant which fills its Roche lobe and more
massive primary is a MS star and does not fill its Roche lobe. With using
depth of minima, period and primary spectrum we could classify AO~Cas
as hot semidetached algol (SH), but secondary spectrum Sp$_2$ is somewhat
earlier as compared with other SH systems. 

\noindent\textbf{AD~Cap}\\
This system was classified as cold semidetached binary (SC). But according to data of \citet{Pojmanski2002}  depth of the primary and secondary minima  equal to $0^m.33$. However the lower limit of A$_1$ is larger than $0^m.5$ for SC systems. Also the value of period of AD~Cap is smaller as compared with other SC. \citet{Antonopoulou1987} have pointed out that AD~Cap is chromospheric active system of RS CVn type. The depth of both minima and period are appropriate to DR class, but spectra are not. The spectra of AD~Cap is later than upper limit for DR class.

As a result
we have compiled a
list of about 39 such systems, which are attractive for additional observations
and further investigations. All of these systems have (reliably known) values of parameters that 
distinguish them from other classified systems and lead to contradictory classification.

\subsection{Extreme and unusual systems}
When our algorithm have been applied to CEV, we found a number of systems with unusual parameters: periods, spectra, depth of minima, eccentricity. These systems do not fall at any of the classes that we use. Such systems belong to unusual 
stages of evolution (for example pre-MS systems) and are rare.
Some of them are listed below, the full list of unusual systems contains 54 binaries.

\noindent\textbf{Post common envelope binaries}\\   
We have found that several detached systems with
white dwarf have very large depth of the primary
minimum: UU~Sge, RR~Cae, GK~Vir, QS~Vir. 
They are all turned out to be post common envelope systems (also known as pre-cataclysmic binaries). 

\noindent\textbf{BY Dra variables}\\
We have found also three detached chromospherically active systems with late-type subdwarf spectra (dK--dM): YY~Gem, CM~Dra and BB~Scl. They all have been classified as DM (detached main sequence binaries) but their periods are shorter as compared to other DM systems.  They all are well known variables of BY~Dra type. 

\noindent\textbf{OW~Gem}\\                 
OW~Gem consists of two supergiants and was classified as DG system but, according to GCVS data and ASAS-3 light curve, it has an extreme value of the phase of secondary minima (0.23P) which points to the high eccentricity of the orbit. We have found it incredible and have checked OW~Gem with literature.  We have found that system consists of two stars with quite different masses. \citet{Terrell2003} have noted that  if the two stars were formed together both cannot
possibly be in the red giant stage. Also significant mass transfer is ruled out by very small
relative radii of stars and the very large orbital
eccentricity.   \citet{Eggleton2002} proposed a suggestion that  OW~Gem is a former triple system with the primary having formed from the merger of a close binary. But the merger product might be
expected to have rapid rotation, and although the primary
star does appear to be rotating faster than the pseudosynchronous rate, it is not unusually rapid. Primary could have undergone a G/K supergiant stage just after the merger and had substantial angular
momentum removed by stellar wind and magnetic braking
during a period of enhanced activity. This hypothesis needs to be tested.

\section{Conclusions}\label{conclusions}
A new version of the Catalogue of Eclipsing Variables (CEV) is
constructed. The catalogue contains
parameters for some 7200 stars. Also we have included to CEV
recently published information about classification of 1352 systems,
and, therefore, it represents the largest list of eclipsing binaries
classified from observations.
The catalogue can be used for classification and parameterization
of known objects, characterization of new populations and
discoveries of unusual objects. 
We have performed the comprehensive analysis of the distribution of 
observable stellar parameters of eclipsing systems of our catalogue 
and have developed an algorithm for the classification of eclipsing variables.

The procedure was applied to all catalogued EBs, and some
  systems remain unclassified. Part of them belong to marginal
  evolutionary classes. Observational data for others are too
  contradictory, and additional observations are needed to
  attribute them to one or another evolutionary class (lists of
  unclassified systems are available upon request).

\section*{Acknowledgements} 
We thank Laurent Corp for his observations of some interesting eclipsing binaries.
This work has been supported by Russian Foundation for Fundamental
Research grants 10-02-00426, 12-02-31904 and 12-07-00528,
by the Federal Science and Innovations
Agency under contract 02.740.11.0247,
by the Presidium RAS program ``Leading Scientific
Schools Support'' 3602.2012.2, and
by the Federal task program ``Research and
operations on priority directions of development of the
science and technology complex of Russia for 2007-2013''
(contract 14.518.11.7064).
This research has made use of the SIMBAD and VizieR databases, 
operated at the Centre de
Donn\'ees astronomiques de Strasbourg,
of the International 
Variable Star Index (VSX) database, operated at AAVSO, Cambridge, 
Massachusetts, USA, and
of NASA's Astrophysics Data System Bibliographic Services.

\bibliographystyle{ceab}
\bibliography{extreme}
\end{document}